# Iteration-Promoting Variable Step Size Least Mean Square Algorithm for Accelerating Adaptive Channel Estimation

Beiyi Liu[†1], Guan Gui[1], Li Xu[1] and Nobuhiro Shimoi[2]

1. Department of Electronics and Information Systems, Akita Prefectural University, Yurihoujo, 015-0055 Japan
2. Department of Machine Intelligence and Systems Engineering, Akita Prefectural University, Yurihoujo, 015-0055 Japan

(Tel: +81-184-27-2241; E-mail: guiguan@akita-pu.ac.jp)

**Abstract:** Invariable step size based least-mean-square error (ISS-LMS) was considered as a very simple adaptive filtering algorithm and hence it has been widely utilized in many applications, such as adaptive channel estimation. It is well known that the convergence speed of ISS-LMS is fixed by the initial step-size. In the channel estimation scenarios, it is very hard to make tradeoff between convergence speed and estimation performance. In this paper, we propose an iteration-promoting variable step size based least-mean-square error (IPVSS-LMS) algorithm to control the convergence speed as well as to improve the estimation performance. Simulation results show that the proposed algorithm can achieve better estimation performance (3dB) than previous ISS-LMS while without sacrificing convergence speed as well as computational complexity.

## 1. INTRODUCTION

Broadband signal transmission is becoming one of the mainstream techniques in the next generation wireless communication systems [1][2]. The channel becomes severely frequency-selective and accurate channel state information (CSI) of such a channel is required for coherent detection (or demodulation). One of the effective approaches is the adaptive channel estimation (ACE) using invariable step size least mean square error (ISS-LMS) algorithm [3], which has low complexity and can be easily implemented at the receiver. In the channel estimation scenarios, step-size of the ACE is the critical parameters to balance the estimation performance and convergence speed in different signal-noise-ratio (SNR) regimes. Hence, ISS-LMS using the only one step-size may be hard to adjust estimation performance and convergence speed under different SNR scenarios. In other words, ISS-LMS using bigger step-size can achieve faster convergence speed but obtain worse estimation performance, and vice versa.

Motivated by the research background, variable step size based least mean square error (VSS-LMS) algorithm is expected to effectively balance estimation performance and convergence speed. In last decade, instantaneous updating error based VSS-LMS algorithms have been proposed [4]–[11]. In contract to ISS-LMS, these already proposed VSS-LMS algorithms can get extra performance gain by adjusting their step-sizes while at the cost of slow convergence speed and/or high computational time. Because the step-size of these methods depend on updating channel estimation error, i.e., step-size is enlarged/reduced if the estimation error big/small. Indeed, the merit of these VSS-LMS algorithms is that the step-size can be adjusted easily in accordance with the updating error. *However, the tracking procedure to the updating error may generate additional computational burden*. In large updating error scenario, the VSS approaches to initial step-size; while in small updating error scenario, the VSS reduces to a parameter which is decided by some threshold. The detailed theory analysis will be given in section 3. Aforementioned theoretical fact implies that step-size can set empirically to reduce itself in somewhat suitable stable range but without increasing needless computation complexity.

In this paper, hence, a simple iteration-promoting variable step-size based VSS-LMS algorithm (IPVSS-LMS) is proposed. On the one hand, the step-size is devised to reduce gradually as increasing number of adaptive iterations. On the other hand, fast convergence speed is also kept by adopting a hard threshold parameter that can terminate the proposed algorithm.

Unlike conventional VSS-LMS algorithms [4]–[11] as well as ISS-LMS [3], the proposed IPVSS-LMS algorithm can efficiently balance updating estimation performance and convergence speed. Hence, the proposed algorithm can achieve lower steady-state mean square error (MSE) estimation performance than previous algorithms [3]–[9] but without scarifying any convergence speed, which is almost the same as the ISS-LMS. The main work of this paper is summarized as follows. Firstly, the iteration-promoting variable step-size is devised for IPVSS-LMS algorithm. The equivalence of different proposed step-sizes is briefly discussed. Secondly, suitable threshold parameter is selected to control the termination of the algorithm's updating. In addition, steady-state MSE performance of the proposed algorithm is also derived. Finally, computer simulation results are given to confirm the effectiveness of the IPVSS-LMS algorithm.

The remainder of the rest paper is organized as follows. A system model is first described and then the drawback of ACE using ISS-LMS algorithm is pointed out in Section 2. In Section 3, iteration-promoting VSS-LMS algorithm is proposed to accelerate the convergence speed of ACE as well as to improve accuracy of the channel estimation. Computer simulation results are presented in Section 4 to show the performance of the proposed algorithm. Finally, we conclude the paper in Section 5.

† Beiyui Liu is the presenter of this paper.

*Notation*: Throughout the paper, matrices and vectors are represented by boldface upper case letters and boldface lower case letters, respectively; the superscripts $(\cdot)^T$, $(\cdot)^H$, $Tr(\cdot)$ and $(\cdot)^{-1}$ denote the transpose, the Hermitian transpose, the trace and the inverse operators, respectively; $E\{\cdot\}$ denotes the expectation operator.

## 2. PROBLME FORMULATION

Consider a baseband-equivalent frequency-selective fading wireless communication system where the channel $\boldsymbol{w}=[w_0,w_1,\cdots,w_{N-1}]^T$ is $N$ dimensional signal vector and each channel tap satisfies random Gaussian distribution as $\mathcal{CN}(0,1)$ . Assume that an input training signal $\boldsymbol{x}(n)$ is used to probe the unknown sparse channel. At the receiver side, the corresponding observed signal $y(n)$ is given by

$$y(n)=\boldsymbol{w}^T\boldsymbol{x}(n)+z(n) \tag{1}$$

where $\boldsymbol{x}(n)=[x(n),x(n-1),\cdots,x(n-N+1)]^T$ denotes the length-$N$ vector of input signal $x(n)$; $z(n)$ is an additive white Gaussian noise (AWGN), which is assumed to be independent with $\boldsymbol{x}(n)$; The objective of ACE is to adaptively estimate the unknown channel vector $\boldsymbol{w}$ using the training signal vector $\boldsymbol{x}(n)$ and the observed signal $y(n)$.

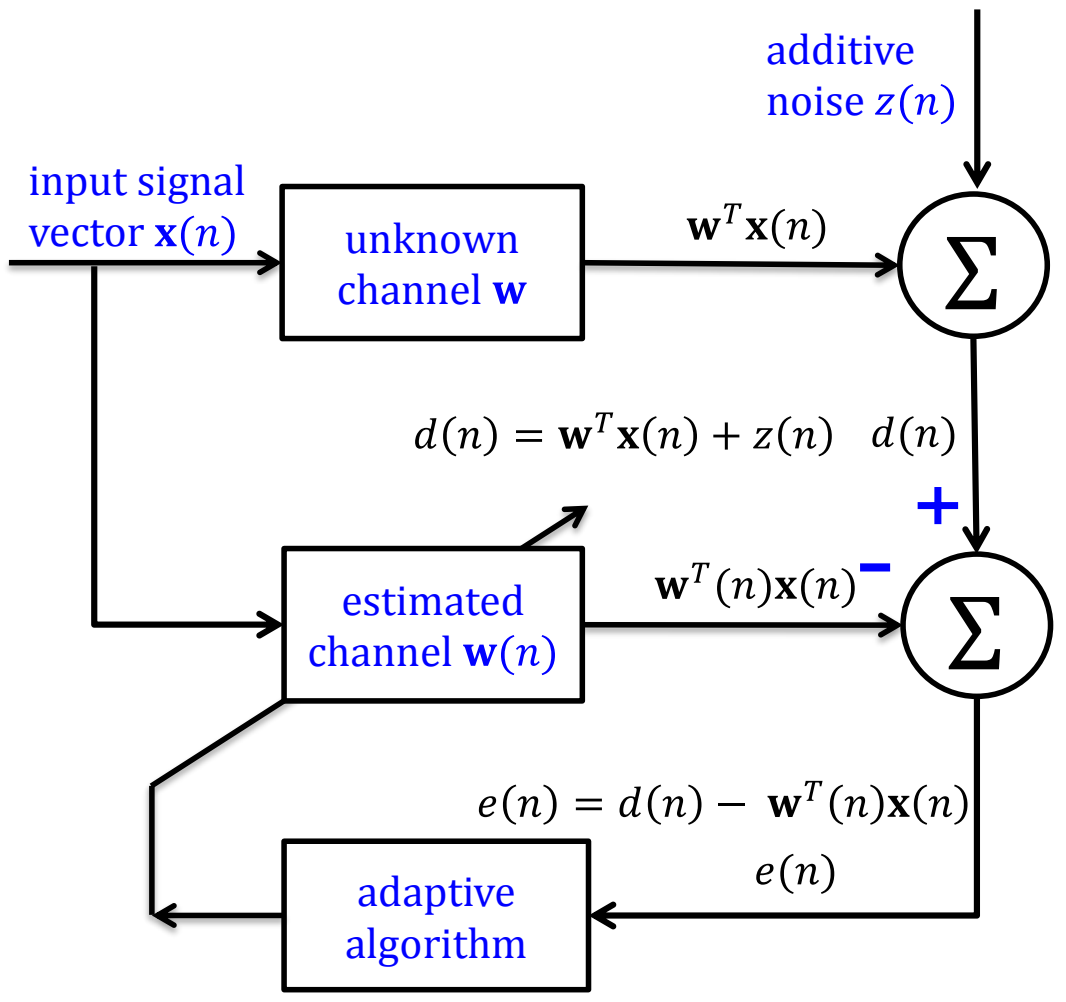

Fig.1 A framework of adaptive channel estimation.

According to (1), standard ISS-LMS based ACE method (see Fig. 1) is overviewed as follows. The cost function of ISS-LMS [3] is constructed as

$$L(n)=(1/2)e^2(n) \tag{2}$$

where $e(n)$ denotes $n$-th update error as

$$\begin{aligned} e(n) &= y(n)-\boldsymbol{w}^T(n)\boldsymbol{x}(n) \\ &= z(n)+(\boldsymbol{w}-\boldsymbol{w}(n))^T\boldsymbol{x}(n) \\ &= z(n)+\boldsymbol{v}^T(n)\boldsymbol{x}(n) \end{aligned} \tag{3}$$

where $\boldsymbol{v}(n)=\boldsymbol{w}-\boldsymbol{w}(n)$ denotes channel estimation error and $\boldsymbol{w}(n)$ represents n-th channel estimator. According to (2), the update equation is derived as

$$\begin{aligned} \boldsymbol{w}(n+1) &= \boldsymbol{w}(n)+\mu\frac{\partial L(n)}{\partial \boldsymbol{w}(n)} \\ &= \boldsymbol{w}(n)+\mu e(n)\boldsymbol{x}(n) \end{aligned} \tag{4}$$

where $\mu\in(0,1/\lambda_{\max})$ is the ISS; $\lambda_{\max}$ is the maximum eigenvalue of covariance matrix $\boldsymbol{R}_{xx}=E[\boldsymbol{x}(n)\boldsymbol{x}^T(n)]$ . Under the independence assumption, in [12], the steady-state MSE of ISS-LMS estimator $\boldsymbol{w}(n)$ is derived as

$$\begin{aligned} \kappa_\mu(\infty) &= \lim_{n\to\infty} E\left\{\left[(\boldsymbol{w}-\boldsymbol{w}(n))^T\boldsymbol{x}(n)\right]^2\right\} \\ &= \frac{Tr\left[\boldsymbol{R}_{xx}(I-\mu\boldsymbol{R}_{xx})^{-1}\right]\sigma_n^2}{2-Tr\left[\boldsymbol{R}_{xx}(I-\mu\boldsymbol{R}_{xx})^{-1}\right]} \\ &\geq \frac{\lambda_{\max}\sigma_n^2}{2-3\mu\sigma_n^2} \end{aligned} \tag{5}$$

where $\sigma_n^2$ denotes the variance of the noise variable of the $z(t)$. One can easily find that the lower bound of steady-state MSE performance depends highly on the step-size $\mu$. If $\mu$ approach to zero, then (5) can be further written as

$$\begin{aligned} \lim_{\mu\to 0}\kappa_\mu(\infty) &\geq \lim_{\mu\to 0}\frac{\lambda_{\max}\sigma_n^2}{2-3\mu\sigma_n^2} \\ &\geq \frac{\lambda_{\max}\sigma_n^2}{2}. \end{aligned} \tag{6}$$

Hence, ISS-LMS using smaller step-size $\mu$ may achieve lower steady-state MSE performance. The idea case is that ISS-LMS adopts big step-size to achieve fast convergence speed while utilizes small step-size to get lower MSE performance. It is difficult to change the step-size adaptively during the gradient descend. This practical problem motivates us to develop iteration-promoting VSS-LMS algorithm in next section.

## 3. PROPOSED IPVSS-LMS ALGORITHM

According to (4), the update equation of IPVSS-LMS algorithm can be written as

$$\begin{aligned} \boldsymbol{w}(n+1) &= \boldsymbol{w}(n) - \mu(n)\frac{\partial L(n)}{\partial \boldsymbol{w}(n)} \\ &= \boldsymbol{w}(n) + \mu(n)e(n)\boldsymbol{x}(n) \end{aligned} \tag{7}$$

where IPVSS $\mu(n)$ is devised as

$$\mu(n) = \begin{cases} \phi, & \text{if } \mu(n) \le \phi \\ \mu/n, & \text{if } \mu(n) > \phi \end{cases} \tag{8}$$

where $\phi$ denotes hard threshold parameter to ensure convergence if IPVSS $\mu(n)$ is enough small. Hence, step-size in (8) can realize two functions: 1) in large estimation error scenario, IPVSS can accelerate the convergence speed; and 2) in quasi-steady-state scenario, the estimation performance is accurate enough and then hard threshold $\phi$ can ensure fast convergence speed as well.

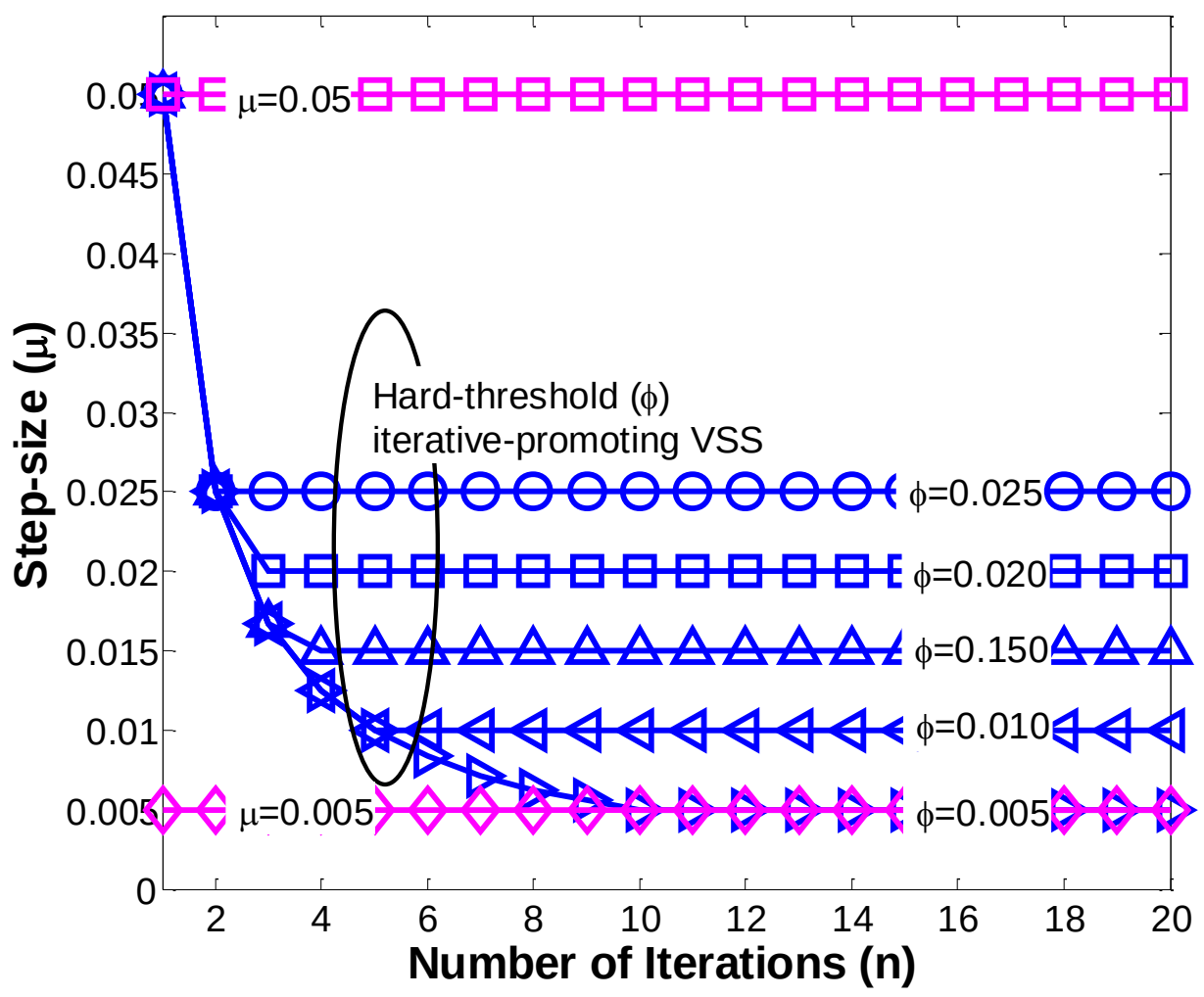

Fig. 2 IPVSS v.s. hard-threshold.

For a better understanding, we briefly discuss the differences of the proposed IPVSS-LMS with other VSS-based adaptive filtering algorithms, e.g., VSS-LMS [6][13] and normalized least mean fourth algorithm (NLMF) [10][11]. The step-size of VSS-LMS [6][13] is decided by

$$\mu_{vss}(n+1) = \mu_0 \cdot \frac{\boldsymbol{p}^T(n+1)\boldsymbol{p}(n+1)}{\boldsymbol{p}^T(n+1)\boldsymbol{p}(n+1)+C}, \tag{9}$$

where $C$ is a positive threshold parameter which is related to $\sigma_n^2 \mathrm{Tr}\{[\boldsymbol{x}(n)\boldsymbol{x}^T(n)]^{-1}\}$ and can be written as $C \sim \mathcal{O}(1/\mathrm{SNR})$, where SNR is the received signal noise ratio (SNR). According to Eq. (9), the range of VSS is given by $\mu_{vss}(n+1) \in (0, \mu_0)$, where $\mu_0$ is the maximal step-size. To the adaptive algorithm stability, the maximal step-size is less than 2 [3]. It is worth mentioning that $\boldsymbol{p}(n)$ in Eq. (9) is defined as

$$\boldsymbol{p}(n+1) = \eta \boldsymbol{p}(n) + (1-\eta)\frac{\boldsymbol{x}(n)e(n)}{\|\boldsymbol{x}(n)\|_2^2}, \tag{10}$$

where $\eta \in [0,1)$ is the smoothing factor for controlling the VSS and estimation error. If $\eta = 0$, then (9) can be rewritten as

$$\begin{aligned} \mu_{vss}(n+1) &= \mu_0 \cdot \frac{\boldsymbol{p}^T(n+1)\boldsymbol{p}(n+1)}{\boldsymbol{p}^T(n+1)\boldsymbol{p}(n+1)+C} \\ &= \mu_0 \cdot \frac{e^2(n)}{e^2(n)+C\|\boldsymbol{x}(n)\|_2^2}. \end{aligned} \tag{11}$$

In the initial updating, if $e^2(n) \gg \|\boldsymbol{x}(n)\|_2^2$, then (11) can approach to

$$\begin{aligned} \lim_{e^2(n)\gg\|\boldsymbol{x}(n)\|_2^2} v_{vss}(n+1) &= \lim_{e^2(n)\gg\|\boldsymbol{x}(n)\|_2^2} \frac{\mu_0 e^2(n)}{e^2(n)+C\|\boldsymbol{x}(n)\|_2^2} \\ &= \mu_0, \end{aligned} \tag{12}$$

which is equivalent to IPVSS in the case of $n=1$. However, the update estimation error and additional computational complexity which is still required for updating VSS in (9). Unlike it, IPVSS in (7) is only calculated by the iteration numbers and a threshold. Hence, the proposed algorithm is very comparably simple as standard ISS-LMS algorithm [3]. Similarly, the proposed IPVSS-LMS is equivalent to step-size of NLMF [10][11] in the case of large estimation error (e.g., $e^2(n) \gg \|\boldsymbol{x}(n)\|_2^2$). That is

$$\begin{aligned} \lim_{e^2(n)\gg\|\boldsymbol{x}(n)\|_2^2} \mu_f &= \lim_{e^2(n)\gg\|\boldsymbol{x}(n)\|_2^2} \frac{\mu_0 e^2(n)}{e^2(n)+\|\boldsymbol{x}(n)\|_2^2} \\ &= \mu_0. \end{aligned} \tag{13}$$

According to above discussions, both (12) and (13) insinuate that step-size could be set as big to speed up fast convergence speed in the initial updating stage where $e^2(n) \gg \|\boldsymbol{x}(n)\|_2^2$. Hence, our proposed IPVSS is logical and then steady-state MSE is derived as

$$\begin{aligned} \kappa_\mu(\infty) &= \lim_{n\to\infty} E\left\{\left[(\boldsymbol{w}-\boldsymbol{w}(n))^T \boldsymbol{x}(n)\right]^2\right\} \\ &= \frac{Tr\left[\boldsymbol{R}_{xx}\left(I-\mu(n)\boldsymbol{R}_{xx}\right)^{-1}\right]\sigma_n^2}{2-Tr\left[\boldsymbol{R}_{xx}\left(I-\mu(n)\boldsymbol{R}_{xx}\right)^{-1}\right]} \\ &\ge \frac{\lambda_{\max}\sigma_n^2}{2-3\mu(n)\sigma_n^2} \ge \frac{\lambda_{\max}\sigma_n^2}{2-3\mu_{\min}\sigma_n^2} \end{aligned} \tag{14}$$

In (14), the initial step-size $\mu(n)$ is the same as ISS-LMS while it decreases as the iteration time ($n$), as shown in Fig. 2. In the first stage, the main demand is fast convergence speed which is decide by the iteration-promoting step-size ( $\mu/n$ ). In the second stage, the key performance indicator is steady-state MSE performance which is decided by the hard threshold or the minimum step-size ( $\mu_{\min}$ ). Hence, the proposed algorithm can be applied in ACE to obtain fast convergence speed as well as to achieve lower steady-state MSE performance.

## 4. COMPUTER SIMULATION

To validate the effectiveness of the proposed method, steady-state MSE standard is adopted to compare the results via $M = 1000$ independent Monte-Carlo runs. Here, the MSE metric is defined as

$$Average\ MSE\{\boldsymbol{w}(n)\} = \frac{1}{M}\sum_{i=1}^{M}\left\|\boldsymbol{w}-\boldsymbol{w}_i(n)\right\|_2^2 \quad (15)$$

The received SNR is defined as $P_0/\sigma_n^2$, where $P_0$ is the received power of the pseudo-random noise (PN)-sequence for training signal. Parameters for computer simulation are given in Tab. I.

Tab. I. Simulation parameters.

| Parameters | Values |
|---|---|
| Training signal | Pseudorandom Binary sequence |
| Channel length | $N = 16$ |
| Distribution of each channel coefficient | Random Gaussian $\mathcal{CN}(0,1)$ |
| Received SNR | 0dB～20dB |
| Step-size of standard ISS-LMS | 0.05 and 0.005 |
| Step-size of VSS-LMS | 0.05 |
| Hard threshold for VSS-LMS | 0.005, 0.01, 0.015, 0.02, 0.025 |

### 4.1. Performance comparisons of IPVSS-LMS in different SNR regimes

Average MSE performance of the proposed method is evaluated in Figs. 3-7 under different SNR regimes, i.e. 0dB~20dB. To confirm the effectiveness of the proposed method, they are compared with standard ISS-LMS algorithm [3]. In the case of different SNR regimes, the proposed algorithm always achieves better performance with respect to average MSE while faster convergence speed with respect to iteration times than ISS-LMS. Since the proposed IPVSS-LMS algorithm adopts the iteration-promoting step-size (i.e., $\mu(n)$) to achieve better performance while utilizing a minimum step-size (threshold) to ensure convergence efficiently. Let us take the Fig. 6 for example to further illustrate the advantages of the proposed algorithm. Two performance curves of standard LMS are depicted by using two step-sizes (0.005 and 0.05) as for benchmarks.

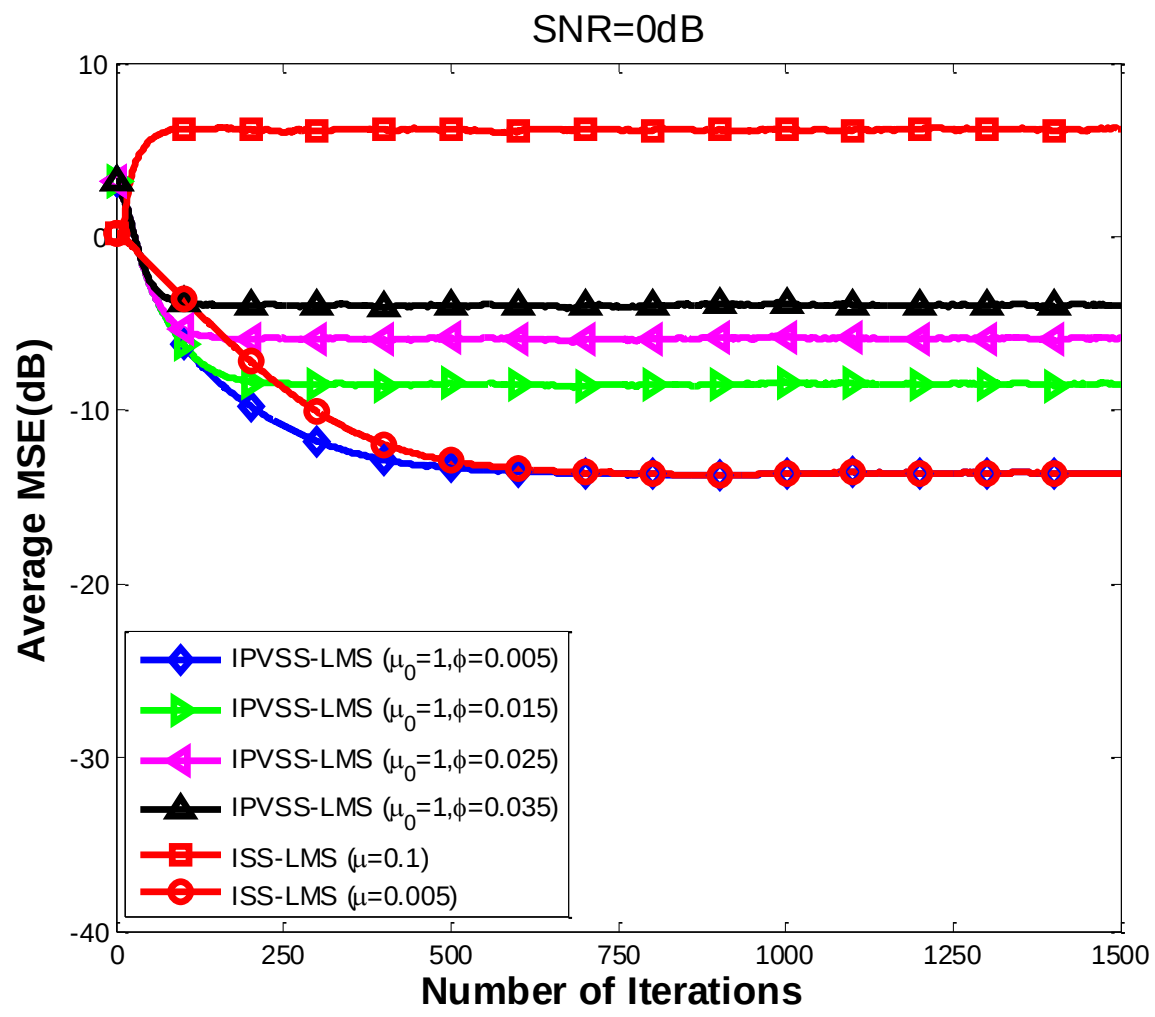

Fig. 3 MSE performance comparisons (SNR=0dB).

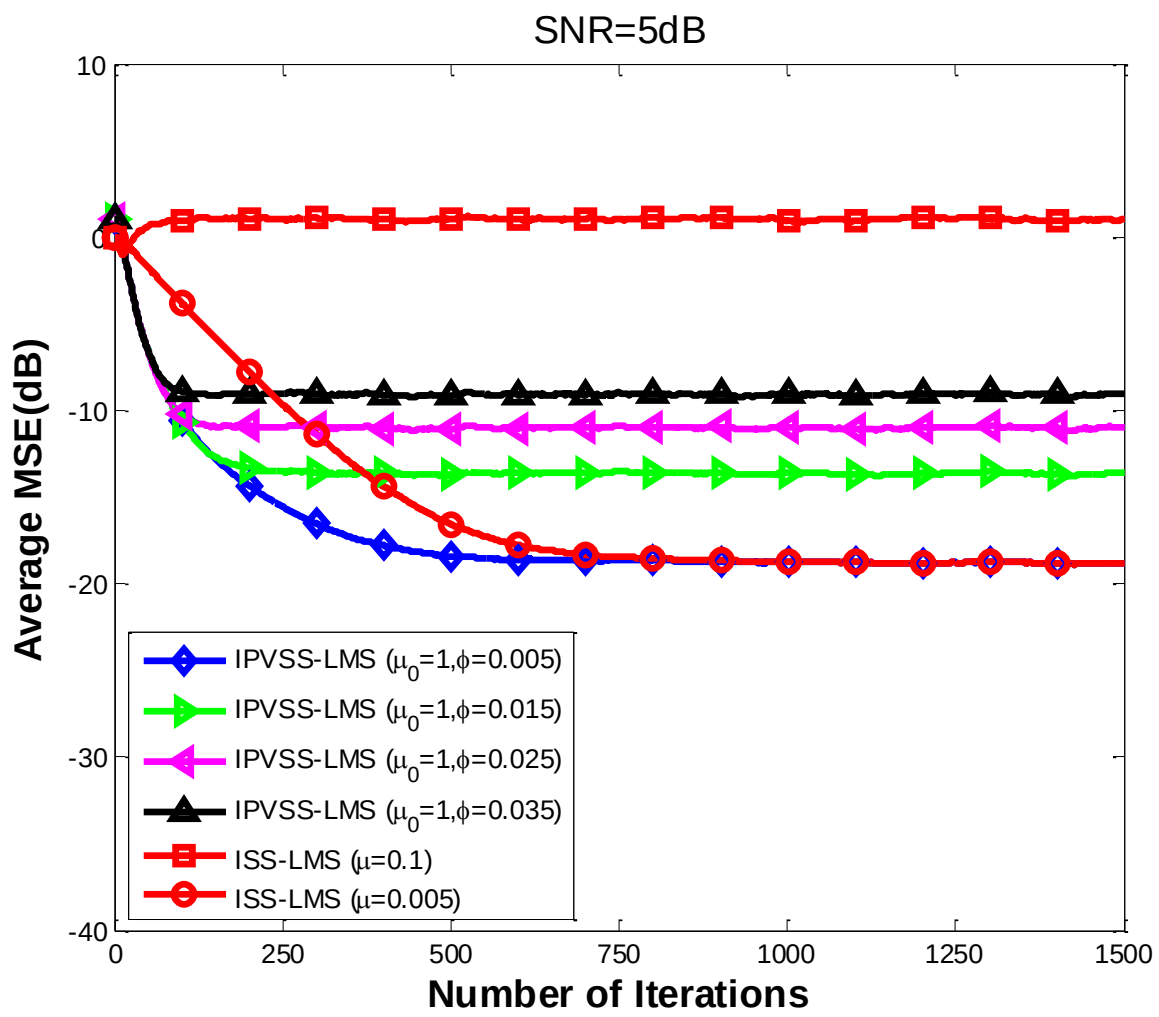

Fig. 4 MSE performance comparisons (SNR=5dB).

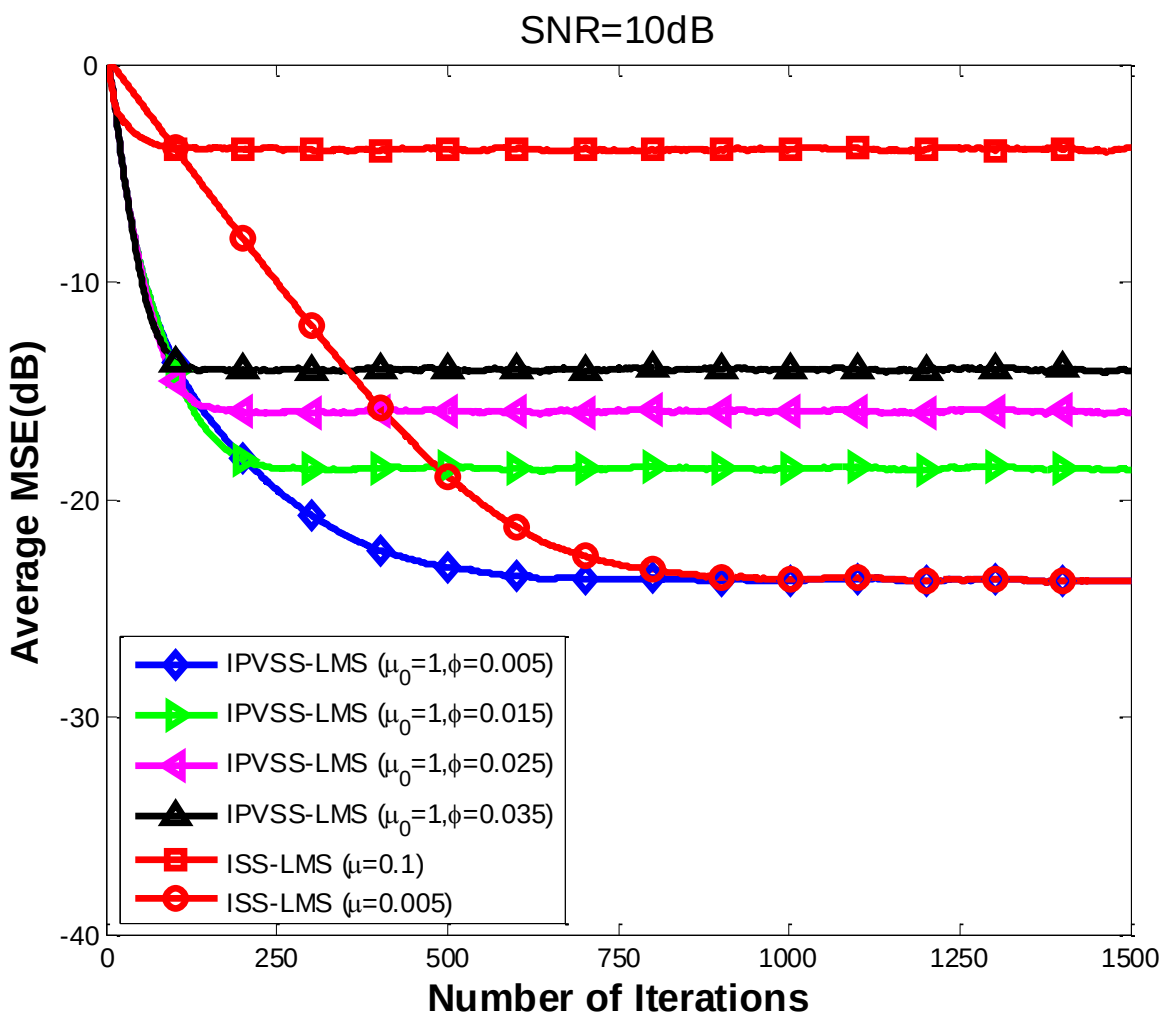

Fig. 5 MSE performance comparisons (SNR=10dB).

On the one hand, if the threshold is set as $\phi = 0.005$, the steady-state performance curve of VSS-LMS is the same as ISS-LMS using step-size $\mu = 0.005$. It is obviously observed that the convergence speed of IPVSS-LMS is faster than ISS-LMS. On the other hand, if the threshold is set as $\phi = 0.025$, the convergence speed of VSS-LMS is almost same as ISS-LMS using step-size $\mu = 0.05$. While the steady-state MSE performance curve is lower than ISS-LMS.

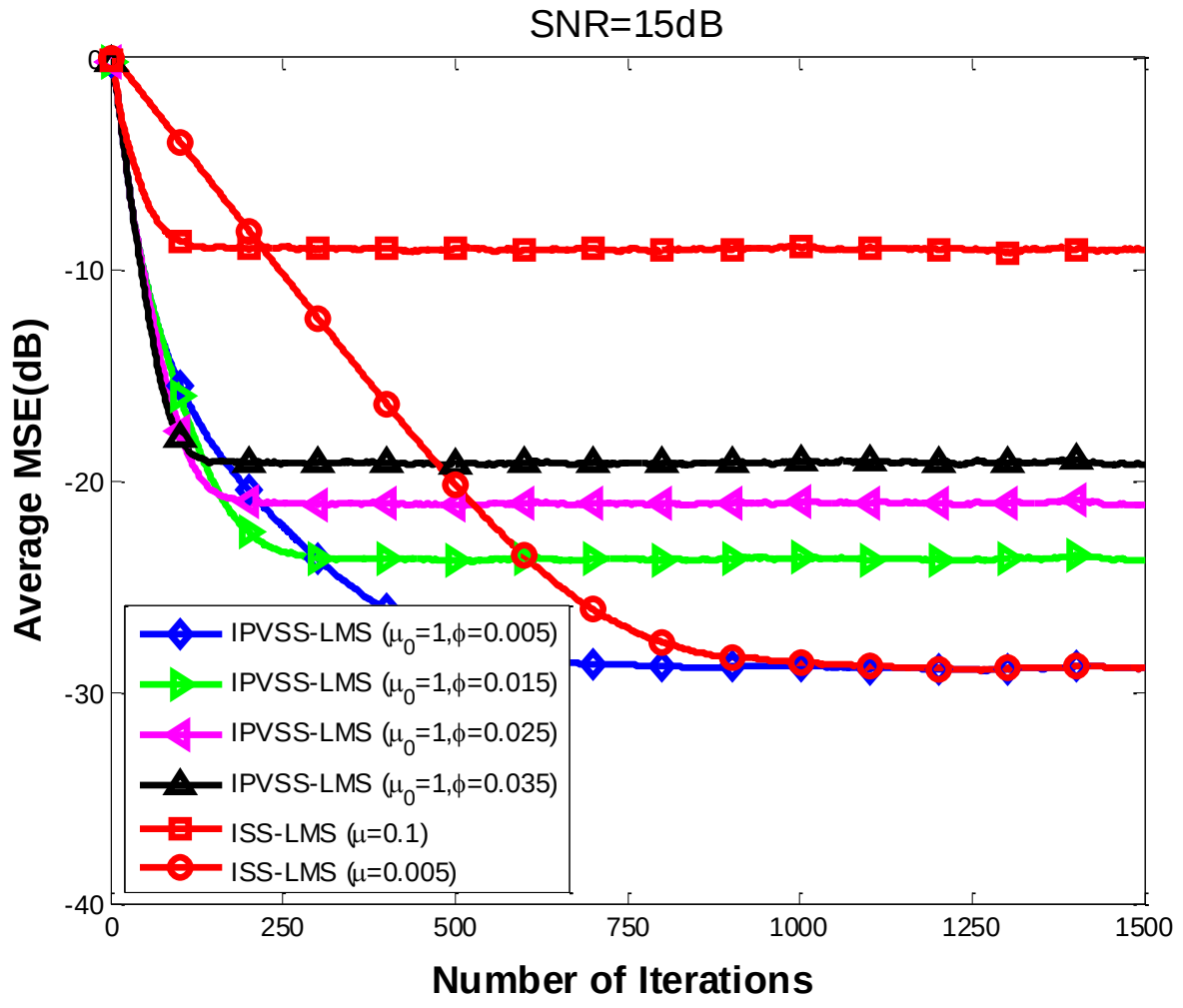

Fig. 6 MSE performance comparisons (SNR=15dB).

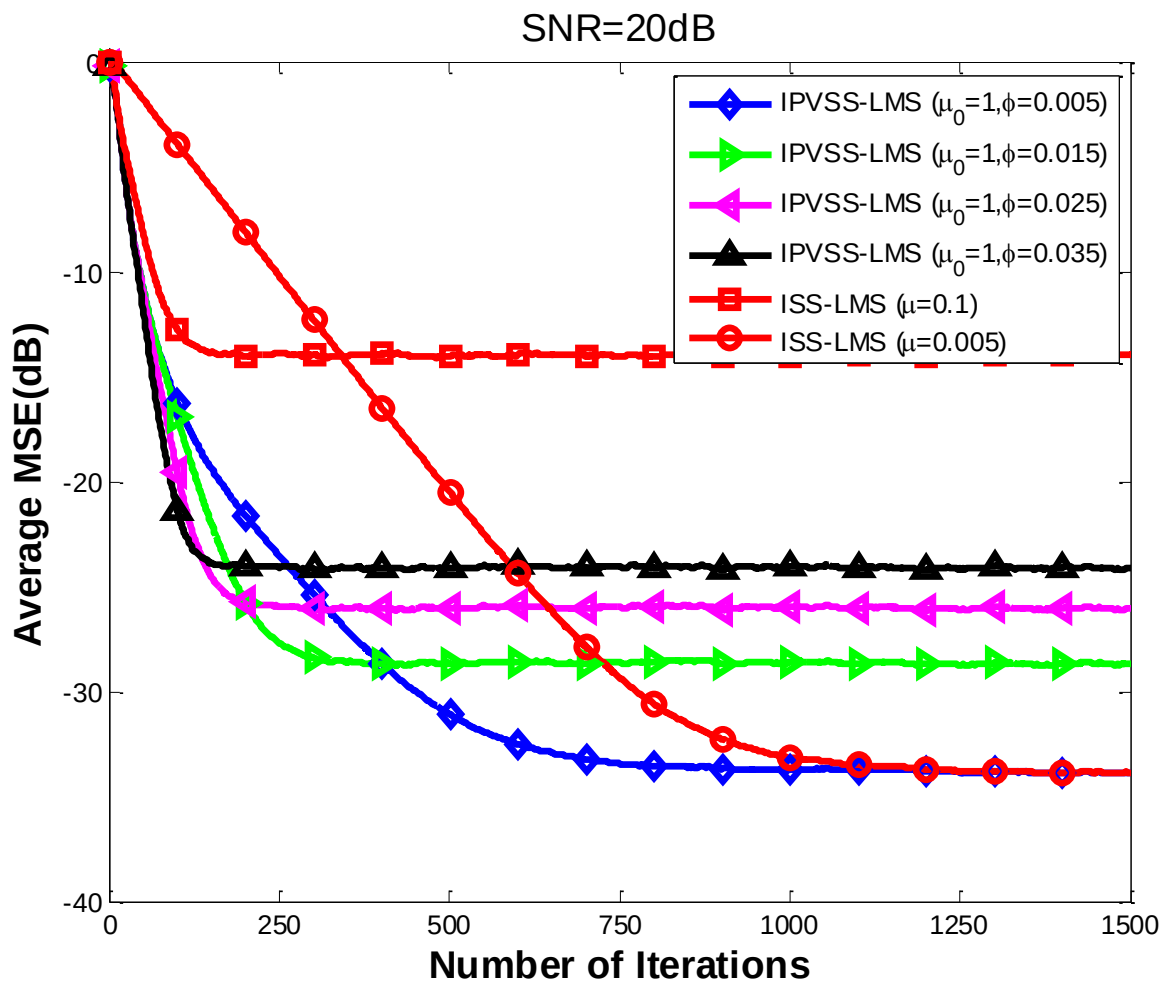

Fig. 7 MSE performance comparisons (SNR=20dB).

### 4.2. Comparing IPVSS-LMS with other methods in MSE and computation complexity

Without loss of generality, the average MSE performance curves of IPVSS-LMS and VSS-LMS are depicted as in Fig. 8.where the hard threshold is set as $\phi = 0.035$. On the one hand, IPVSS-LMS can achieve faster convergence speed than conventional one. On the second hand, IPVSS-LMS can keep almost the same MSE performance as conventional one. To further confirm the proposed method, its computational complexity is compared with the VSS-LMS [6]. It is worth noting that the computational complexity is the arithmetic complexity, which includes additions and multiplications. The complexities of the IPVSS-LMS algorithm and conventional one are shown in Table II. From Table II, we can see that the computational complexity of our IPVSS-LMS algorithm is lower than conventional VSS-LMS which is due to the calculation of updating step-size.

Tab. II. Computational complexity.

| Algorithm | Multiplications | Additions |
|---|---|---|
| ISS-LMS [3] | $2N$ | $2N+1$ |
| VSS-LMS [6] | $6N+6$ | $5N-1$ |
| IPVSS-LMS | $2N+1$ | $2N+1$ |

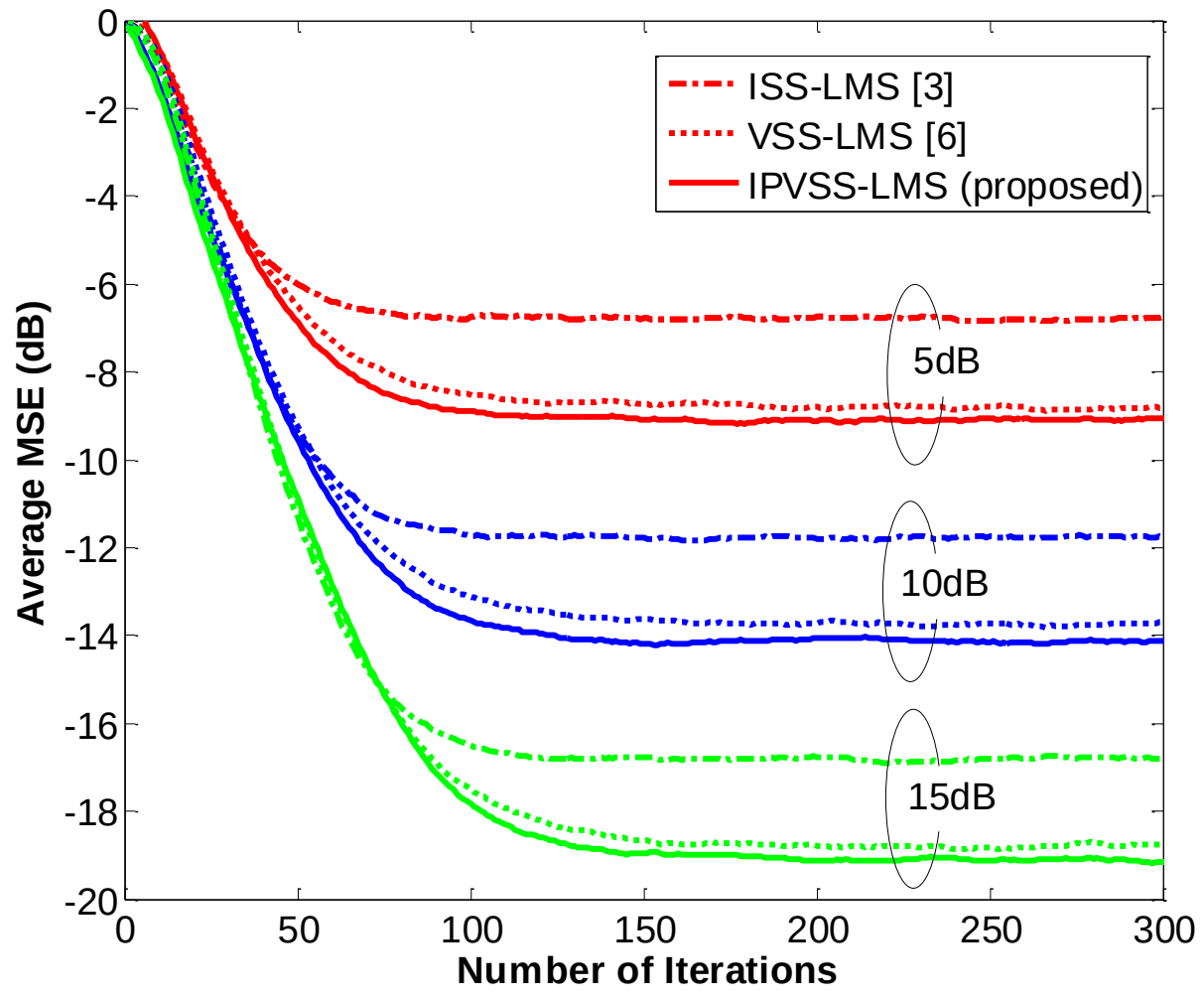

Fig. 8 MSE performance comparisons.

## 5. CONCLUSIONS

This paper proposed an IPVSS-LMS algorithm to balance between the convergence speed and steady-state MSE performance. We first derived the update equation of the proposed IPVSS-LMS algorithm and devised suitable threshold to control step-size efficiently. In addition, we confirmed that equivalence of the three type algorithms, i.e., IPVSS-LMS, VSS-LMS and NLMF algorithms, in big estimation error scenarios. Simulation results show that the proposed algorithm can achieve better estimation performance than previous ISS-LMS while without increasing convergence speed as computational complexity. In future work, steady-state performance analysis of the proposed IPVSS-LMS will be studied. In addition, this proposed method will be applied in multiple-antenna wireless communications systems.